\documentclass{article}
\usepackage{graphicx}
\begin{document}
\title{How to prepare quantum states that follow classical paths}
\author{Pouria Pedram\thanks{pouria.pedram@gmail.com}\\
 {\small Plasma Physics Research Center, Science and Research
Branch, Islamic Azad University, Tehran, Iran}}
\date{\today}
\maketitle \baselineskip 24pt

\begin{abstract}
We present an alternative quantization procedure for the
one-dimensional non-relativistic quantum mechanics. We show that,
for the case of a free particle and a particle in a box, the
complete classical and quantum correspondence can be obtained using
this formulation. The resulting wave packets do not disperse and
strongly peak on the classical paths. Moreover, for the case of the
free particle, they satisfy minimum uncertainty relation.
\end{abstract}


\textit{Pacs}: {03.65.-w}\\

\section{Introduction}
Schr\"odinger equation $H\psi=i\hbar\dot{\psi}$ is the main equation
of non-relativistic quantum physics which can be obtained upon the
quantization procedure $\vec{p}\rightarrow-i\hbar\vec{\nabla}$ and
$H\rightarrow i\hbar\partial_t$ in the Hamiltonian formulation of
quantum mechanics \cite{Ervin2,Ervin3}. In fact, this equation
determines the time evolution of the particle's wave function.
Often, we are interested to find solutions in such a way that they
follow the classical trajectories without dispersion and peak on
them. But, the construction of such kind of wave packets, except for
the case of the simple harmonic oscillator, is not an easy task. For
instance, for the case of a free particle, the initial wave function
disperses quickly as the particle moves.

The problem of classical and quantum correspondence has
attracted much attention in the literature \cite{Bolivar}. These
efforts have begun by Schr\"odinger \cite{Ervin} and followed by
others in the context of the coherent states. These wave packets are
specific kind of quantum states that describe a maximal kind of
coherence and a classical kind of behavior \cite{ref}. In quantum
physics, we can construct such kind of quantum states by the
superposition of the energy eigenstates which would peak around the
classical trajectories. But, in most cases, the wave packets do not
maintain their shape and eventually disperse.

One possible way to solve this problem is using an alternative
quantization method. In fact, we need to change the differential
structure of the Schr\"odinger equation which has a parabolic
form. It has been shown that in the context of quantum cosmology, the
hyperbolic nature of its main equation gives us the possibility of
complete classical and quantum correspondence
\cite{pedram1,pedram2}. So, the resulting wave packets strongly peak
on the classical trajectories and never disperse.

Here, we first consider a one-dimensional model in the presence of
an external general potential. Then, we use an alternative classical
picture which, after quantization, results in a hyperbolic
differential equation. For the case of a free particle and a
particle in a box, we solve this equation and construct wave packets
using appropriate initial conditions. We show that these wave
packets follow the classical paths and strongly peak on them in the
whole configuration space.

\section{The model}
Let us consider a harmonic oscillator as a simple example. In the
classical domain, this model has a well-known solution
$u(t)=A\cos(\omega t+\delta_1)$. Note that, time explicitly appears
in this solution which parameterize the temporal behavior of $u$. To
eliminate the explicit presence of $t$, we can use an another
general solution with the same total energy but different phase
\emph{i.e.} $v(t)=A\cos(\omega t+\delta_2)$. Now, we can write the
variable $u$ in terms of $v$ instead of $t$, namely
\begin{eqnarray}
u=\cos(\Delta)v\pm \sin(\Delta)\sqrt{A^2-v^2},
\end{eqnarray}
where $\Delta=\delta_1-\delta_2$ and $-A\leq u,v\leq A$. So, in
general, the trajectory is an ellipse which for $\Delta=\pi/2$
represents a circle. This example shows that we can always
parameterize the solution in terms of another one which has the same
energy but with arbitrary phase shift. Moreover, for each model, the
trajectories are unique due to the specific form of the potential.

Since the both solutions have a same energy and obey a same equation
of motion we have
\begin{eqnarray}
\left\{
  \begin{array}{l}
  \frac{\displaystyle p^2_u}{\displaystyle2m}+V(u)=E,\\ \\
  \frac{\displaystyle p^2_v}{\displaystyle2m}+V(v)=E.
\end{array}
\right.
\end{eqnarray}
Since the right hand sides are equal, we obtain
\begin{eqnarray}
\mathcal{F}\equiv \frac{\displaystyle p^2_u}{\displaystyle2m}-
\frac{\displaystyle p^2_v}{\displaystyle2m}+V(u)-V(v)=0.
\end{eqnarray}
In the quantum mechanical domain, we need to quantize above
equation. The operator $\mathcal{F}$ can be obtained upon
quantization procedure $p_u\rightarrow
-i\hbar\frac{\displaystyle\partial}{\displaystyle\partial u}$ and
$p_{v}\rightarrow
-i\hbar\frac{\displaystyle\partial}{\displaystyle\partial v}$. So,
we demand that $\mathcal{F}$ annihilate the wave function
\emph{i.e.}
\begin{eqnarray}
\mathcal{F}\Psi(u,v)=0,
\end{eqnarray}
or, equivalently
\begin{equation}\label{wd}
\left\{- \frac{\displaystyle \hbar^2}{\displaystyle2m}
\frac{\partial^2}{\partial u^2}+\frac{\displaystyle
\hbar^2}{\displaystyle2m}\frac{\partial^2}{\partial
v^2}+V(u)-V(v)\right\}\, \Psi(u,v)=0.
\end{equation}
Therefore, Eq.~(\ref{wd}) is an alternative quantum mechanical
equation which has a different structure with respect to the
Schr\"odinger equation, but both explain the same physics.

Although there is no intrinsic preference between these two
pictures, the later has some additional advantages. This is due to
the hyperbolic form of Eq.~(\ref{wd}) which gives us freedom for
choosing the initial wave function and its initial slope. On the
other hand, the solutions of a hyperbolic equation are usually
highly oscillatory. Since the oscillation around the classical paths
is not acceptable, we need to choose appropriate initial conditions
to guarantee the classical--quantum correspondence.

In the next section, to show the method, we consider the case of a
free particle. For this case, the trajectories in both $u-t$ and
$u-v$ planes are straight lines. Since Eq.~(\ref{wd}) admits
solutions which never disperse and strongly peak on the classical
paths, we show that the classical and quantum correspondence arises
naturally from this formulation.

\section{Free particle}
For the case of the free particle ($V=0$), we have
\begin{eqnarray}
\left\{
  \begin{array}{l}
  u(t)=\beta_u t+u_0,\\ \\
  v(t)=\beta_v t+v_0,
\end{array}
\right.
\end{eqnarray}
where $\beta$ is the particle's velocity. Since both solutions have
a same total energy, we also have $|\beta_u|=|\beta_v|$ which
results in $u=\pm v+u_0\mp v_0$. So, the trajectories are straight
lines with unit absolute slope.

\begin{figure}
\centerline{\begin{tabular}{cc}
\includegraphics[width=6cm]{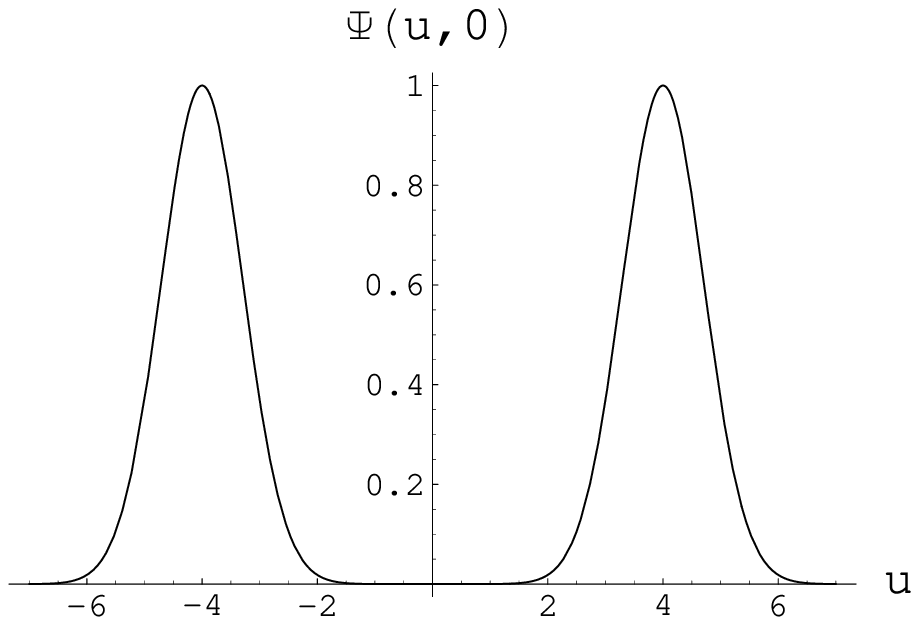}
 &
\includegraphics[width=6cm]{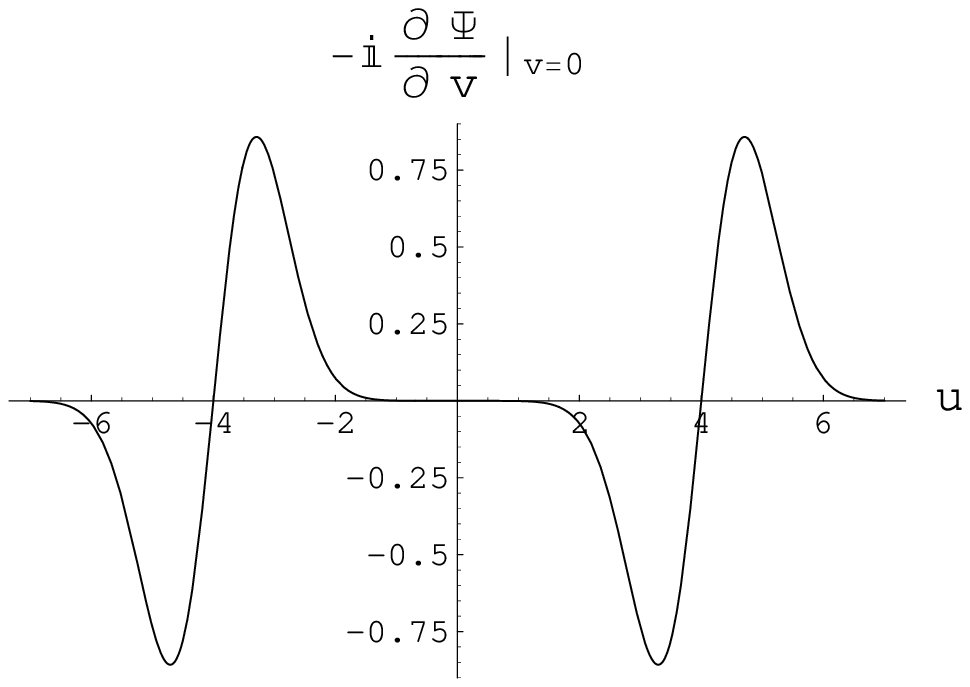}
\end{tabular}}
\caption{The initial wave function (left) and the initial derivative
of the wave function (right) for $d=4$ and $\alpha=1$.} \label{fig1}
\end{figure}

\begin{figure}
\centerline{\begin{tabular}{ccc}
\includegraphics[width=6cm]{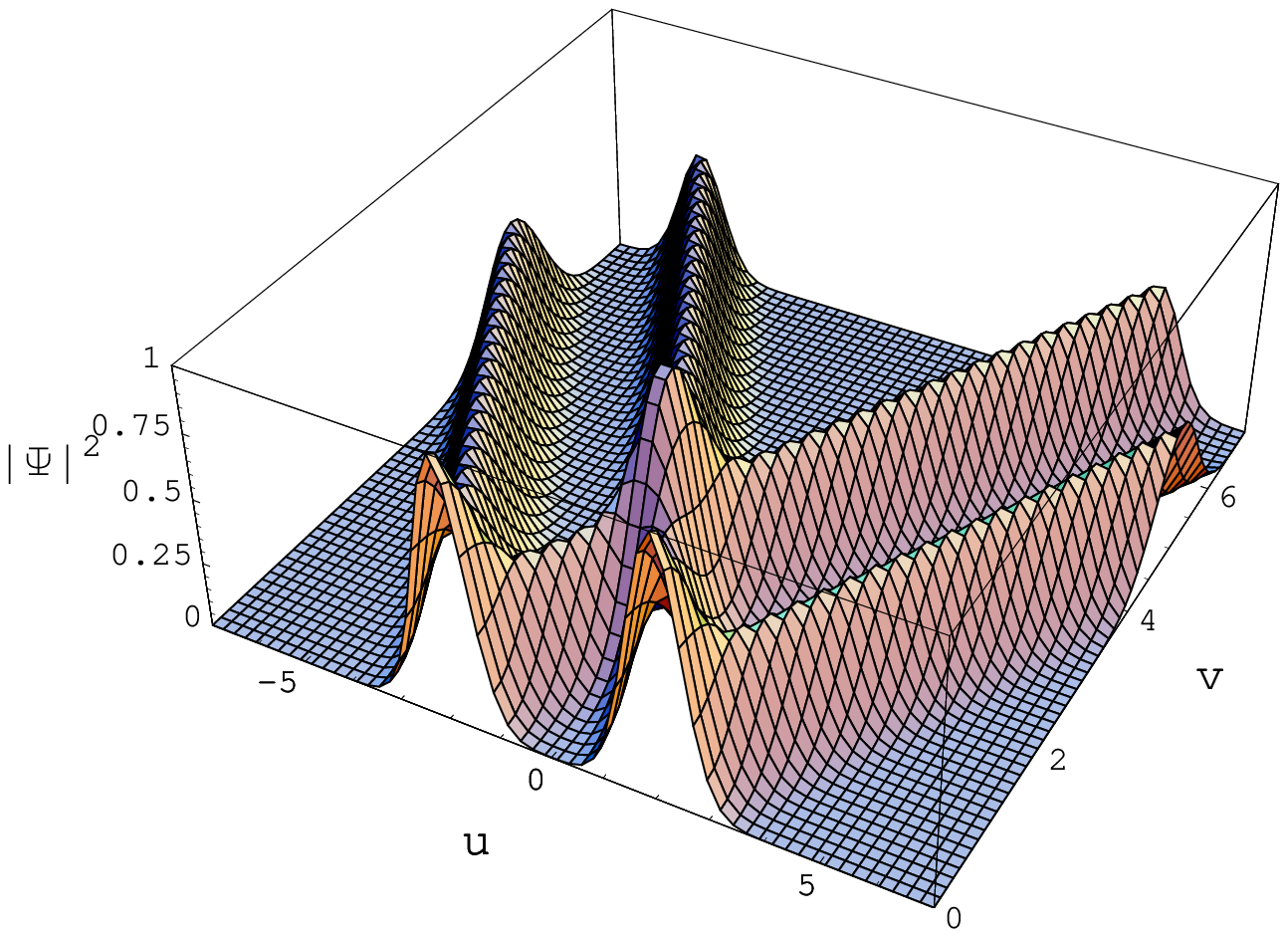}
 &
\includegraphics[width=6cm]{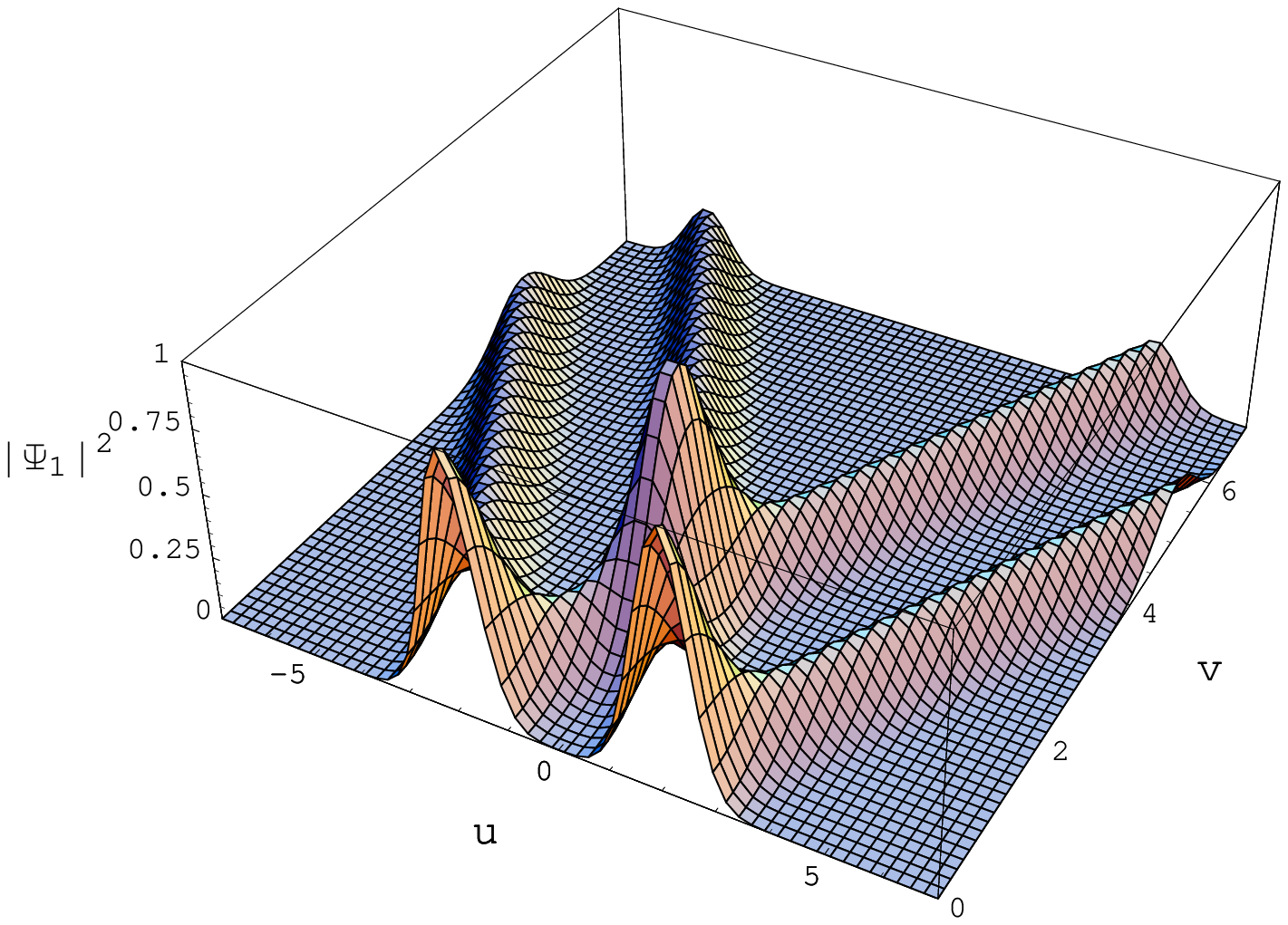}
 &
\includegraphics[width=6cm]{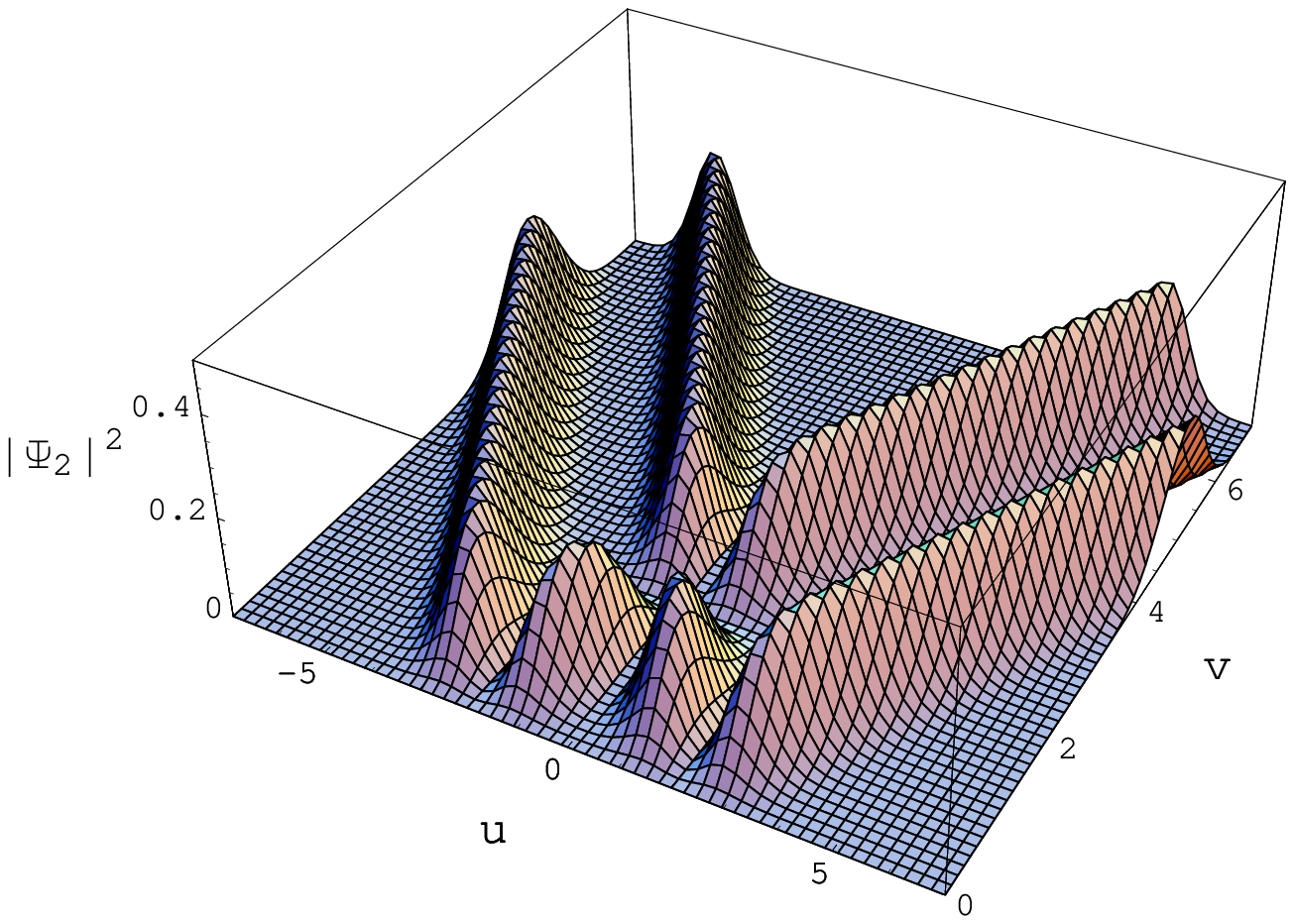}
\end{tabular}}
\centerline{\begin{tabular}{ccc}
\includegraphics[width=4cm]{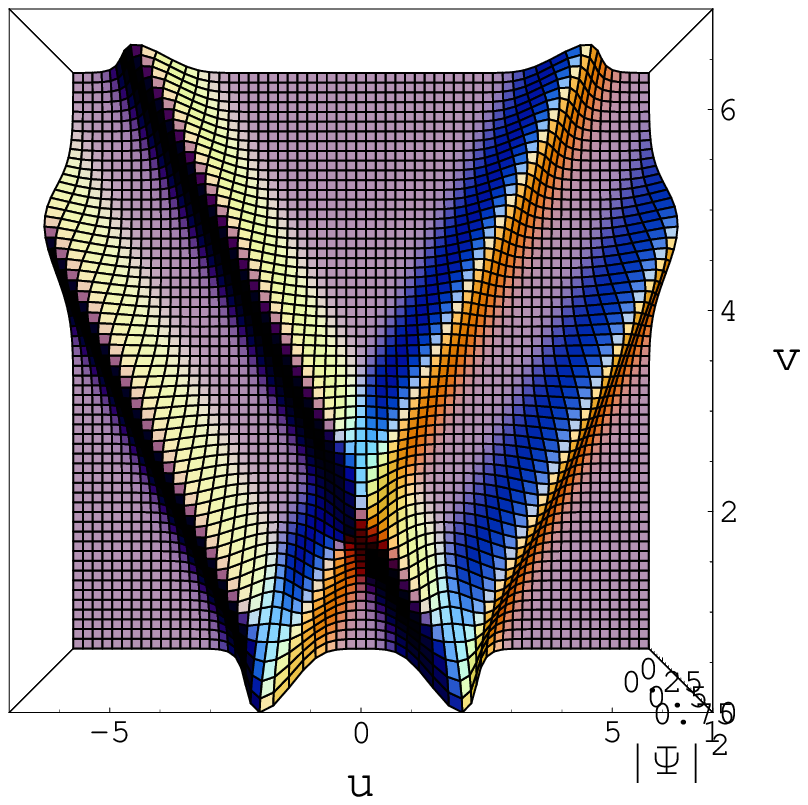}
 &
\includegraphics[width=4cm]{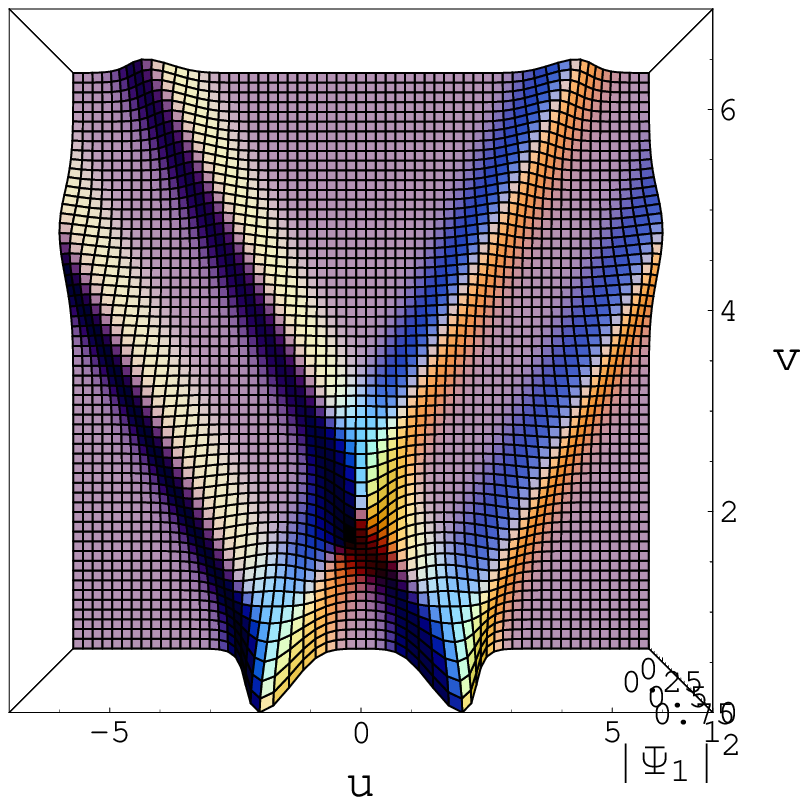}
 &
\includegraphics[width=4cm]{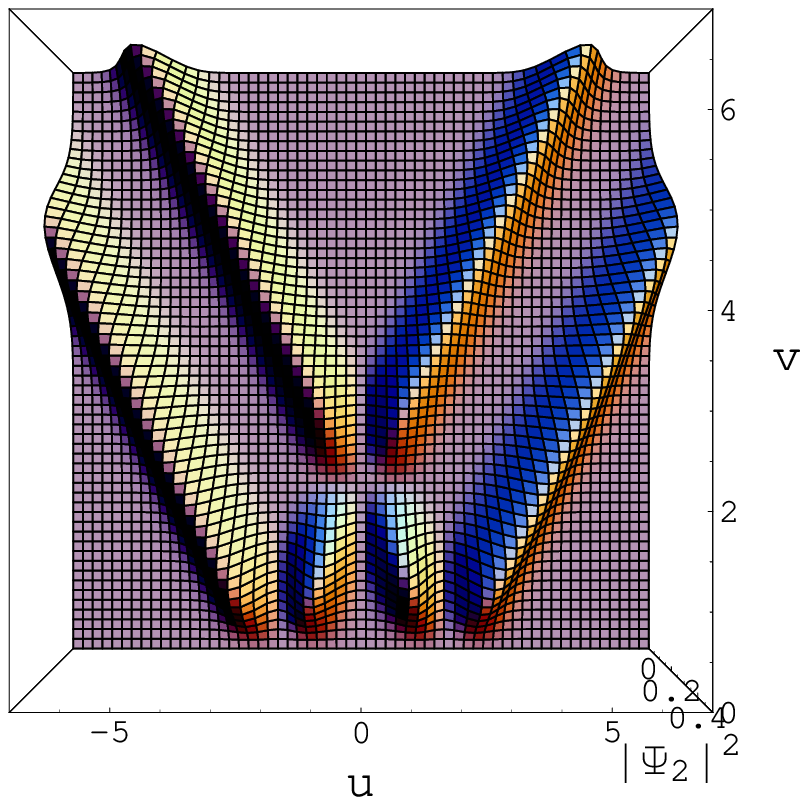}
\end{tabular}}
\caption{Up: the square of the wave packet $|\Psi(u,v)|^2$ (left),
$\left[\mbox{Re}[\Psi(u,v)]\right]^2$ (middle), and
$[\mbox{Im}[\Psi(u,v)]]^2$ (right) for $d=2$ and $\alpha=1$, Down:
the respective upper view.} \label{fig2}
\end{figure}

It is straightforward to check that, for $V=0$, $\cos(ku)\cos(kv)$
and $\sin(ku)\sin(kv)$ are the eigenfunctions of Eq.~(\ref{wd}).
Therefore, the general solution is
\begin{eqnarray}\label{psi}
\Psi(u,v)&=&\frac{1}{\sqrt{2\pi}}
\int_{-\infty}^{\infty}\left[A(k)\cos(ku)\cos(kv)\right.\nonumber \\
&+&\left. i B(k)\sin(ku)\sin(kv)\right]\,dk.
\end{eqnarray}
To find the complete form of the wave packet, we need to specify the
coefficients $A(k)$ and $B(k)$. These coefficients can be determined
from the initial form of the wave packet at $v=0$
\begin{eqnarray}
\Psi(u,0)=\frac{1}{\sqrt{2\pi}}\int_{-\infty}^{\infty} A(k)
\cos(ku)\,d k,\\
\frac{\partial\Psi(u,v)}{\partial
v}\bigg|_{v=0}=\frac{i}{\sqrt{2\pi}}\int_{-\infty}^{\infty} B(k)
\sin(ku)\,k\,d k.
\end{eqnarray}
It is obvious that the presence of $B(k)$ dose not have any effect
on the form of the initial wave function but it is responsible for
the slope of the wave function at $v=0$, and vice versa for $A(k)$.
Moreover, a complete description of the problem would include the
specification of both of these quantities. On the other hand, since
we are interested to construct wave packets with classical
properties, we need to assume a specific relationship between these
coefficients. The prescription is that these coefficients have the
same functional form \cite{pedram1,pedram2} \textit{i.e.}
\begin{eqnarray}\label{eqcanonicalslope}
 A(k)=B(k),
 \end{eqnarray}
which results in the following wave packet
\begin{eqnarray}\label{psi2}
\Psi(u,v)&=&\frac{1}{\sqrt{2\pi}}\int_{-\infty}^{\infty}A(k)[
\cos(ku)\cos(kv)\nonumber \\ &+&i \sin(ku)\sin(kv)]\,dk.
\end{eqnarray}
To completely determine the wave packet, we should specify $A(k)$.
We can find $A(k)$ by choosing an appropriate initial wave function
such as two Gaussians at $u=\pm d$
\begin{equation}\label{ini}
\Psi(u,0)=e^{-\alpha\left( u-d \right)^2} +e^{-\alpha\left( u+d
\right)^2 }.
\end{equation}
This choice of initial condition, using inverse fourier transform,
is related to
\begin{equation}\label{coef}
A(k)= \sqrt{\frac{2}{\alpha} }\,e^
     {-\frac{k^2}{ 4\alpha} }\cos(kd),
\end{equation}
which gives
\begin{eqnarray}\label{psi3}
\Psi(u,v)&=&\frac{1-i}{2}\left(e^{-\alpha ( u + v +d)^2}+e^{-\alpha
( u + v -d)^2}\right. \nonumber \\ &+& \left. ie^{-\alpha ( u - v
+d)^2}+ie^{-\alpha ( u - v -d)^2}\right).
\end{eqnarray}
Figure \ref{fig1} shows the initial wave function and its initial
slope for $d=4$ and $\alpha=1$. Since, from Bohmian interpretation
of the quantum mechanics, the initial derivative of the imaginary
part of the wave packet corresponds to the initial classical
velocity, the presence of a nonzero and appropriate form of $B(k)$
can be justified. The plot of the initial slope of the wave packet
contains a negative and a positive peaks which correspond to a
incoming or outgoing particle, respectively. Figure \ref{fig2} shows
the resulting wave packet for $d=2$ and $\alpha=1$. As it can be
seen from the figure, this wave packet never disperses and strongly
peaks on the classical trajectory. Moreover, the height of the crest
of the wave packet, which from WKB approximation corresponds to the
inverse velocity of the particle, is constant along the classical
path. This behavior is in complete agreement with the classical
picture. However, the square of the wave packet increases at the
intersection points of the trajectories which correctly corresponds
to the probability  of two possible particle's directions of motion.
Note that our choice of the expansion coefficients (\ref{coef})
corresponds to a free particle with a positive or negative momentum
which is located initially at $u=\pm d$.

\section{Particle in a box}
\begin{figure}
\centerline{\begin{tabular}{cc}
\includegraphics[width=7cm]{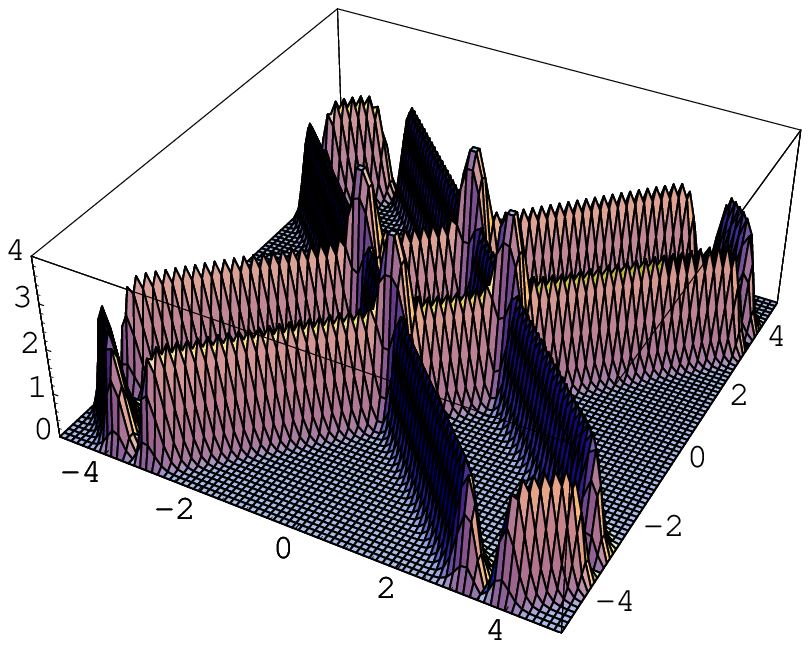}
 &
\includegraphics[width=5cm]{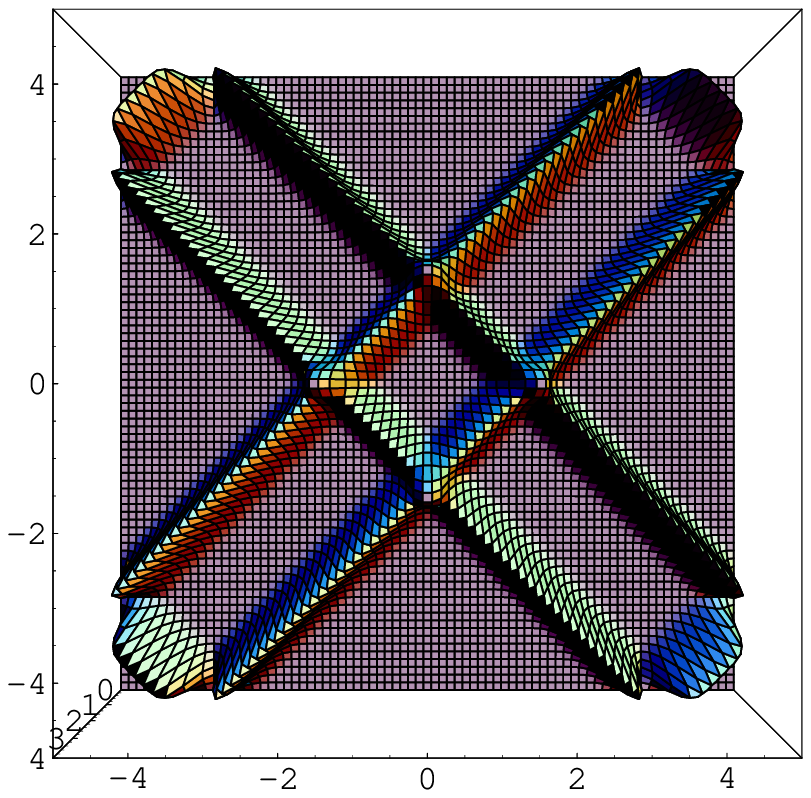}
\end{tabular}}
\caption{The square of the wave packet $|\Psi(u,v)|^2$ (left), and
the respective upper view (right) for $d=1.5$ and $\alpha=5$.} \label{figwell}
\end{figure}

For a particle in a box, we have
\begin{equation}
V(q)= \left\{
\begin{array}{ll}
0\qquad\qquad -\frac{L}{2}< q<\frac{L}{2},\\ \\
\infty\qquad\qquad \mbox{otherwise},
   \end{array}\displaystyle
   \right.
\end{equation}
where $q$ stands for $u$ or $v$. For this case, the trajectories
also are straight lines with unit absolute slope. Moreover, this
model has well-known orthonormal even and odd eigenfunctions
\begin{equation}
\psi_n(q)=\left\{
\begin{array}{ll}
\sqrt{\frac{2}{L}}\cos(\frac{n \pi q}{L}),\qquad\qquad n=1,3,5,...,\\ \\
\sqrt{\frac{2}{L}}\sin(\frac{n \pi q}{L}),\qquad\qquad n=2,4,6,....
\end{array}\displaystyle
\right.
\end{equation}
Now, using the exact form of the eigenstates, we find the following
wave packet
\begin{eqnarray}
\Psi(u,v)&=&\sum_{n=1,3,5,...}A(n)\cos(\frac{n \pi
u}{L})\cos(\frac{n \pi v}{L})\nonumber \\ &+&i\sum_{n=2,4,6,...}A(n)
\sin(\frac{n \pi u}{L})\sin(\frac{n \pi v}{L}).
\end{eqnarray}
If we demand that this solution satisfies the initial condition of
Eq.~(\ref{ini}), we obtain the following form of the expansion
coefficients
\begin{eqnarray}\nonumber
A(n)&=&\frac{-ie^{\frac{n\pi \left( n\pi  +  4i \alpha d L \right)
}{4\alpha L^2 }}}{L{\sqrt{\alpha /\pi }}}\left[-
\mbox{Erfi}(\frac{n\pi  +
2i\alpha L \left( d - L \right) }{2{\sqrt{\alpha }}L})\right.\\
\nonumber &+& e^{\frac{2i dn\pi }{L}} \left(\mbox{Erfi}(\frac{n\pi
- 2i \alpha L\left( d - L \right) }{2{\sqrt{\alpha }}L})\right.\\
\nonumber &-& \left.
  \mbox{Erfi}(\frac{n\pi  - 2i \alpha L\left( d + L \right)}{2{\sqrt{\alpha
  }}L})\right)\\&+&\left.
  \mbox{Erfi}(\frac{n\pi  + 2i \alpha L\left( d + L \right)
 }{2{\sqrt{\alpha }}L})\right],
\end{eqnarray}
where $\mbox{Erfi}(x)$ is the imaginary error function
$\mbox{Erfi}(x)\equiv-i\,\mbox{Erf}(ix)$. Figure \ref{figwell} shows
the wave packet for $d=1.5$ and $\alpha=5$. Classically, the
particle is free inside the well and moves with positive or negative
initial velocity from any arbitrary position in the range $
-L/2<u,v<L/2$. As it can seen from the figure, the wave packet
follows the classical path and strongly peaks on it which is in
complete agreement with the classical scenario. In fact, the value
of $d$ determines the initial classical position and the presence of
two rectangles with opposite directions indicates the two possible
directions of motion. Since the height of the crest of the wave
packet is constant along the classical trajectory, the probability
of finding the particle is constant along the classical path. On the
other hand, since the classical velocity of the particle for this
case is a constant of motion, the classical probability of finding
the particle is also constant along its trajectory. So, the
classical probability completely coincides with the quantum
mechanical probability.

\section{Classical limit, notion of time, uncertainty principle, and the conservation law}
Now, we can define the classical wave packet which is nearly zero
beyond the classical path as follows:
\begin{eqnarray}
\Psi_{cl}(u,v)\equiv\lim_{\alpha\rightarrow\infty}\Psi(u,v).
\end{eqnarray}
This wave packet has the desired property $\Psi_{cl}(u,v)\ne0$ at
$\{u=u_{cl},v=v_{cl}\}$ and $\Psi_{cl}(u,v)\simeq0$ elsewhere. In
fact, the wave packet does not disperse even for large values of
$\alpha$. This is in contrast with the solutions of the
Schr\"odinger equation which usually disperses quickly as the particle moves.

The notion of time does not appear explicitly in our main equation
(\ref{wd}). However, as we have shown, the results depend on time in
an implicit manner. We also encounter this phenomenon in the
context of quantum cosmology where its main equation, which is the
Wheeler-DeWitt equation, does not contain time. Moreover, it is a
hyperbolic differential equation and in particular conditions can be
written in the form of Eq.~(\ref{wd}) \cite{pedram1}. Therefore, our
approach can be considered as a bridge between relativistic and
non--relativistic quantum mechanics. On the other hand, we can find
the notion of time using the Bohmian interpretation of quantum
mechanics, namely
\begin{eqnarray}
p_{\mu}=\partial_{\mu}S,
\end{eqnarray}
where $\Psi=R\exp(iS)$. So, we can obtain the time evolution of each
variable using this interpretation. Although this definition of time
is not genuine, at the classical limit ($\alpha\rightarrow\infty$)
it will coincide with the classical time as desired.

At this point, we encounter an important question: do these
wave packets satisfy the Heisenberg uncertainty relation? For the
case of a free particle, using the explicit form of the wave packet
(\ref{psi3}), we can check the uncertainty principle, for instance
at $v=0$. At this point, the uncertainties in $u$ and $p_u$ take the
following form
\begin{eqnarray}\nonumber
(\Delta u)^2&=&\frac{1}{4\alpha}+\frac{d^2}{2}\left(1+\tanh(
d^2\alpha)\right)\\ &-&\left(\frac{\sqrt{\frac{2}{\pi
\alpha}}+d^2e^{2
d^2\alpha}\mbox{Erf}(\sqrt{2\alpha}d^2)}{1+e^{2d^2\alpha }}
\right)^2,\\
(\Delta
p_u)^2&=&\alpha\hbar^2\left[1-\frac{4\alpha}{1+e^{2d^2\alpha}}\left(d^2\alpha-\frac{2/\pi}{1+e^{2d^2\alpha
}} \right)\right],
\end{eqnarray}
where the integration is over $\{0,\infty\}$. Since the initial wave
function contains two well-separated Gaussians located at $\pm d$
($\alpha d^2> 1$), we have $(\Delta u)^2\simeq \frac{1}{4\alpha}$
and $(\Delta p_u)^2\simeq \alpha\hbar^2$ which results in $(\Delta
u)^2(\Delta p_u)^2\simeq \hbar^2/4$. So, similar to the coherent
states, the initial form of the wave packet satisfies the minimized
uncertainty relation for all values of $d$ and $\alpha$ subjected to
$\alpha d^2> 1$. Since the wave packet preserves its shape during
the motion, we expect that this result also holds for other values
of $v$.

Note that, in this formulation, the total probability at fixed $u$
or $v$ is not a conserved quantity. In fact, the wave packets which
are solutions of the Eq.~(\ref{wd}) satisfy the following
conservation law
\begin{eqnarray}
\partial_{\mu}J_{\mu}=0,
\end{eqnarray}
where the probability current is defined as
\begin{eqnarray}
J_{\mu}=\frac{\displaystyle
\hbar^2}{\displaystyle2m}(\Psi^*\partial_{\mu}\Psi-\Psi\partial_{\mu}\Psi^*).
\end{eqnarray}
Therefore, the probability interpretation of Eq.~(\ref{wd}) is
similar to the Klein--Gordon equation.

\section{Conclusions}
In this paper, we have presented an alternative quantization
procedure which results in a hyperbolic differential equation. We
showed that, for the case of a free particle and a particle in a
box, the structure of the underlying quantum mechanical equation and
appropriate initial conditions led to complete classical--quantum
correspondence. The wave packets never dispersed and followed the
classical trajectories in the whole configuration space and strongly
peaked on them.

\end{document}